\documentstyle[aps,twocolumn]{revtex}

\begin{document}
\twocolumn[\hsize\textwidth\columnwidth\hsize\csname@twocolumnfalse\endcsname

\title
{Maximal Violation of Bell's Inequalities for Continuous Variable Systems}

\author{Zeng-Bing Chen$^1$, Jian-Wei Pan$^{1,2}$, Guang Hou$^1$, and Yong-De Zhang$^{3,1}$}
\address
{$^1$Department of Modern Physics, University of Science and Technology of China,
Hefei, Anhui 230027, P.R. China}
\address
{$^2$Institut f\"ur Experimentalphysik, Universit\"at Wien, 
Boltzmanngasse 5, 1090 Wien, Austria}
\address
{$^3$CCAST (World Laboratory), P.O. Box 8730, Beijing 100080, P.R. China}
\date{\today}
\maketitle 

\begin{abstract}

\
We generalize Bell's inequalities to biparty systems with continuous 
quantum variables. This is achieved by introducing the Bell operator in perfect 
analogy to the usual spin-$1/2$ systems. It is then demonstrated that 
two-mode squeezed vacuum states display quantum nonlocality by using 
the generalized Bell operator. In particular, the original 
Einstein-Podolsky-Rosen states, which are the limiting case 
of the two-mode squeezed vacuum states, can maximally violate Bell's 
inequality due to Clauser, Horne, Shimony and Holt. The experimental
aspect of our scheme is briefly considered.

\


\

\

\end{abstract}]

In their famous paper \cite{EPR}, Einstein, Podolsky and Rosen (EPR)
introduced two striking aspects of quantum mechanics into physics: quantum
entanglement and quantum nonlocality. The relationship between them has then
been a source of great theoretical interest. These fundamental issues play
an essential role in the modern understanding of quantum phenomena. However,
further studies of quantum nonlocality and entanglement, especially those
providing quantitative tests of quantum mechanics versus local realism in
the form of Bell's inequalities \cite{Bell,CHSH,Bell-book}, used mainly
Bohm's version \cite{Bohm} of the EPR entangled states instead of the
original EPR states with continuous degrees of freedom. In recent years, the
later has attracted much attention. The preparation of the EPR-type states
for photons was investigated both theoretically \cite{Reid,Walls-book} and
experimentally \cite{Ou,Ou-APB}. However as noticed in Refs. \cite{Ou,Ou-APB}%
, the generalization of Bell's inequalities to quantum systems with
continuous variables (CVs) is a challenging issue.

In the burgeoning field of quantum information theory \cite{QIT,Nature}, EPR
entanglement and quantum nonlocality are also of practical importance. The
fascinating nonlocal correlations can be exploited to perform classically
impossible tasks. While most of the concepts in quantum information theory
were initially developed for quantum systems with discrete quantum
variables, quantum information protocols (e.g., quantum teleportation \cite
{BK}, quantum error correction \cite{error}, quantum computation \cite
{computation}, entanglement purification \cite{Duan} and cloning \cite
{cloning}) of CVs have also been proposed very recently.

Quantum nonlocality for position-momentum variables associated with the
original EPR states was analyzed recently \cite{Bell-book,Bell86,Banaszek}.
Using the Wigner function approach \cite{Wigner}, Bell \cite
{Bell-book,Bell86} has argued that the original EPR states will not exhibit
nonlocality, because its Wigner function is positive everywhere, and as such
will allow a hidden variable description of the system. By sharp contrast,
it was demonstrated in recent publications \cite{Banaszek} that the Wigner
function of the two-mode squeezed vacuum states (the ``regularized'' EPR
states), though positive definite, provides a direct evidence of the
nonlocal character of the states. The demonstration is based on the fact
that the Wigner function can be interpreted as a correlation function for
the joint measurement of the parity operator. By making use of the parity of
coherent states as one of the observables, Yurke and Stoler have also
presented a proposal for observing the local realism violation with squeezed
states \cite{Yurke}. Using homodyning with weak coherent fields and photon
counting, a recent experiment \cite{exp} reported the observed violation of
the Bell-type inequalities by the regularized EPR states produced in a
pulsed nondegenerate optical parametric amplifier (NOPA), confirming the
theoretical prediction in Refs. \cite{Banaszek,Grangier}.

There is a crucial point implied in Ref. \cite{Banaszek}: A state does not
have to violate all possible Bell's inequalities to be considered quantum
nonlocal; a given state is nonlocal when it violates {\em any} Bell's
inequality. This point has been also stressed in Ref. \cite{Jeong}. Thus the
degree of quantum nonlocality that we can uncover crucially depends not only
on the given quantum state but also on the ``Bell operator'' \cite{Bell-op}.
In their demonstration of quantum nonlocality of the NOPA states by means of
the phase-space formalism, Banaszek and W\'odkiewicz (BW) used the Bell
operator based on the joint parity measurements \cite{Banaszek}. However, it
still remains to be answered whether or not the original EPR states can
maximally violate Bell's inequalities within the BW formalism. Moreover, the
violation of Bell's inequalities uncovered by BW depends upon the magnitude
of the displacement in phase space, an unsatisfactory feature. Thus the
challenging problem of generalizing Bell's inequalities to quantum systems
with CVs is only {\it partially} solved in Ref. \cite{Banaszek}.

In this paper, we generalize Bell's inequalities to the CV cases for biparty
systems. We then show that the original EPR states, which are the limiting
case of the NOPA states, can maximally violate Bell's inequality due to
Clauser, Horne, Shimony and Holt \cite{CHSH}, called the Bell-CHSH
inequality in the following. In contrast to the BW formalism (using the
phase space approach) and the proposal by Grangier {\it et al}. \cite
{Grangier} (using the homodyne detection scheme), here we show an
interesting and direct analogy between Bell's inequalities for both
discrete-variable and CV cases; the correlation functions to be measured for
observing the violation of the Bell-CHSH inequality are also analogous for
the two cases.

To this end, we need to introduce a Bell operator suitable for the present
purpose. First, let us recall some well known results of the Bell-CHSH
inequality for two-qubit systems (e.g., spin-$1/2$ systems). In the
two-qubit case, the Bell operator reads \cite{Bell-op} 
\begin{eqnarray}
{\cal B}_{qubit} &=&({\bf a}\cdot {\bf \sigma }_1)\otimes ({\bf b}\cdot {\bf %
\sigma }_2)+({\bf a}\cdot {\bf \sigma }_1)\otimes ({\bf b}^{\prime }\cdot 
{\bf \sigma }_2)  \nonumber \\
&&+({\bf a}^{\prime }\cdot {\bf \sigma }_1)\otimes ({\bf b}\cdot {\bf \sigma 
}_2)-({\bf a}^{\prime }\cdot {\bf \sigma }_1)\otimes ({\bf b}^{\prime }\cdot 
{\bf \sigma }_2),  \label{bellspin}
\end{eqnarray}
where ${\bf \sigma }_j$ is the Pauli matrix for the $j$th ($j=1$, $2$)
qubit; ${\bf a}$, ${\bf a}^{\prime }$, ${\bf b}$ and ${\bf b}^{\prime }$ are
four unit three-dimensional vectors. We can easily derive \cite
{Bell-op,Landau} 
\begin{equation}
{\cal B}_{qubit}^2=4I_{2\times 2}+4[({\bf a\times a}^{\prime })\cdot {\bf %
\sigma }_1]\otimes [({\bf b\times b}^{\prime })\cdot {\bf \sigma }_2],
\label{spin2}
\end{equation}
where $I_{2\times 2}$ is the identity operator for the qubit systems. As a
result, the expectation value of ${\cal B}_{qubit}^2$ with respect to a
two-qubit state satisfies $\left\langle {\cal B}_{qubit}^2\right\rangle \leq
4+4=8$, implying that $\left| \left\langle {\cal B}_{qubit}\right\rangle
\right| $ with respect to any two-qubit state is bounded by $2\sqrt{2}$,
known as the Cirel'son bound \cite{MAX}.

Now for a single-mode light field we can introduce the following
``pseudospin'' operators for photons (Perhaps the pseudospin operators have
been introduced in literature somewhere we are unware of) 
\begin{eqnarray}
s_z &=&\sum_{n=0}^\infty \left[ \left| 2n+1\right\rangle \left\langle
2n+1\right| -\left| 2n\right\rangle \left\langle 2n\right| \right] , 
\nonumber  \label{oe} \\
s_{-} &=&\sum_{n=0}^\infty \left| 2n\right\rangle \left\langle 2n+1\right|
=(s_{+})^{\dagger },  \label{s}
\end{eqnarray}
where $\left| n\right\rangle $ are the usual Fock states. The operator $%
s_z=-(-1)^N$ ($N$ is the number operator), where $(-1)^N$ is the parity
operator; $s_{+}$ and $s_{-}$ are the ``parity-flip'' operators. In terms of
the creation ($a^{\dagger }$) and annihilation operators ($a$), $s_{-}$ can
also be written as $s_{-}=\frac{I+(-1)^N}{2\sqrt{N+1}}a$, where $I$ is the
identity operator. It is interesting to note that $\frac 1{\sqrt{N+1}}%
a=e^{i\vartheta }$, with $\vartheta $ known as the Susskind-Glogower phase
operator \cite{Susskind}. We can easily check that 
\begin{equation}
\left[ s_z,s_{\pm }\right] =\pm 2s_{\pm },\;\;\;\;\left[ s_{+},s_{-}\right]
=s_z.  \label{comm}
\end{equation}
The commutation relations in Eq. (\ref{comm}) are identical to those of the
spin-$1/2$ systems. Therefore the pseudospin operator ${\bf \hat s}%
=(s_x,s_y,s_z)$, where $s_x\pm is_y=2s_{\pm }$, can be regarded as a
counterpart of the spin operator ${\bf \sigma }$. It is a kind of spin
operator acting upon the parity space of photons, and thus can be called the
``parity spin'' of photons. The quadrature amplitudes of a single-mode light
field correspond to the usual position and momentum operators, accompanied
with the position-momentum uncertainty. The fact that one can define the
parity spin as in (\ref{s}) might imply a new intrinsic uncertainty for
photons (and other bosons).

Now choosing an arbitrary vector living on the surface of a unit sphere $%
{\bf a}=(\sin \theta _a\cos \varphi _a,\sin \theta _a\sin \varphi _a,\cos
\theta _a)$ [$\theta _a$ ($\varphi _a$) being the polar (azimuthal) angle of 
${\bf a}$], we have 
\begin{equation}
{\bf a}\cdot {\bf \hat s}=s_z\cos \theta _a+\sin \theta _a(e^{i\varphi
_a}s_{-}+e^{-i\varphi _a}s_{+}).  \label{as}
\end{equation}
Analogously, ${\bf a}$ may be interpreted as the ``direction'' along which
we measure the parity spin ${\bf \hat s}$. The commutation relations in Eq. (%
\ref{comm}) lead to 
\begin{equation}
({\bf a}\cdot {\bf \hat s})^2=I.  \label{unit}
\end{equation}
Equation (\ref{unit}) means that the outcome of measurement of the Hermitian
operator ${\bf a}\cdot {\bf \hat s}$ (with eigenvalues $\pm 1$) is $1$ or $%
-1 $. The above observations show that there exists a {\it perfect} analogy
between the CV systems and the usual spin-$1/2$ systems. Thus all types of
Bell's inequalities derived for the latter have their counterpart in the
former.

In particular, for two-mode light fields we define the Bell operator as 
\begin{eqnarray}
{\cal B}_{CHSH} &=&({\bf a}\cdot {\bf \hat s}_1)\otimes ({\bf b}\cdot {\bf 
\hat s}_2)+({\bf a}\cdot {\bf \hat s}_1)\otimes ({\bf b}^{\prime }\cdot {\bf 
\hat s}_2)  \nonumber \\
&&+({\bf a}^{\prime }\cdot {\bf \hat s}_1)\otimes ({\bf b}\cdot {\bf \hat s}%
_2)-({\bf a}^{\prime }\cdot {\bf \hat s}_1)\otimes ({\bf b}^{\prime }\cdot 
{\bf \hat s}_2).  \label{bellop}
\end{eqnarray}
Here ${\bf a}$, ${\bf a}^{\prime }$, ${\bf b}$ and ${\bf b}^{\prime }$ are
four unit vectors as before; ${\bf \hat s}_1$ and ${\bf \hat s}_2$ are
defined as in Eq. (\ref{s}). Then local realistic theories impose the
following Bell-CHSH inequality \cite{CHSH}: 
\begin{equation}
\left| \left\langle {\cal B}_{CHSH}\right\rangle \right| \leq 2,
\label{chsh}
\end{equation}
where $\left\langle {\cal B}_{CHSH}\right\rangle $ is the expectation value
of ${\cal B}_{CHSH}$ with respect to a given quantum state of CVs. Equation (%
\ref{chsh}) represents the Bell-CHSH inequality of quantum systems with CVs.
Interestingly, our generalization of Bell's inequalities to CV systems is
realized via joint measurements on discrete (dichotomic) observable ${\bf 
\hat s}$, in a perfect analogy to the usual joint measurements on spins.
Within this scheme, the {\it correlation function} reads $E({\bf a},{\bf b}%
)=\left\langle ({\bf a}\cdot {\bf \hat s}_1)\otimes ({\bf b}\cdot {\bf \hat s%
}_2)\right\rangle $.

By using $({\bf a}\cdot {\bf \hat s}_1)^2=({\bf b}\cdot {\bf \hat s}_2)^2=(%
{\bf a}^{\prime }\cdot {\bf \hat s}_1)^2=({\bf b}^{\prime }\cdot {\bf \hat s}%
_2)^2=I$ [see Eq. (\ref{unit})] and the commutation relations in Eq. (\ref
{comm}), it can been shown, similarly to the two-qubit case, that 
\begin{eqnarray}
{\cal B}_{CHSH}^2 &=&4I-[{\bf a}\cdot {\bf \hat s}_1,{\bf a}^{\prime }\cdot 
{\bf \hat s}_1]\otimes [{\bf b}\cdot {\bf \hat s}_2,{\bf b}^{\prime }\cdot 
{\bf \hat s}_2]  \nonumber \\
&=&4I+4[({\bf a\times a}^{\prime })\cdot {\bf \hat s}_1]\otimes [({\bf %
b\times b}^{\prime })\cdot {\bf \hat s}_2].  \label{b2}
\end{eqnarray}
Consequently, 
\begin{equation}
\left\langle {\cal B}_{CHSH}^2\right\rangle \leq 4+4=8,  \label{b8}
\end{equation}
which again implies that $\left| \left\langle {\cal B}_{CHSH}\right\rangle
\right| $ with respect to any quantum state of CVs is bounded by $2\sqrt{2}$%
. When $\left| \left\langle {\cal B}_{CHSH}\right\rangle \right| =2\sqrt{2}$
for a given state, we say that the Bell-CHSH inequality (\ref{chsh}) is
maximally violated by the state. In the following we will use the Bell-CHSH
inequality (\ref{chsh}) to uncover nonlocality of the NOPA states as well as
of the original EPR states.

The NOPA process represents a nonlinear interaction of two quantized modes
(denoted by the corresponding annihilation operators $a_1$ and $a_2$) in a
nonlinear medium with a strong classical pump field. In this process the
NOPA can generate the two-mode squeezed vacuum states, i.e., the NOPA states 
\cite{Reid,Walls-book}: 
\begin{equation}
\left| {\rm NOPA}\right\rangle =e^{r(a_1^{\dagger }a_2^{\dagger
}-a_1a_2)}|00\rangle =\sum_{n=0}^\infty \frac{(\tanh r)^n}{\cosh r}%
|nn\rangle ,  \label{nopa}
\end{equation}
where $r>0$ is known as the squeezing parameter and $\left| nn\right\rangle
\equiv \left| n\right\rangle _1\otimes \left| n\right\rangle _2=\frac 1{n!}%
(a_1^{\dagger })^n(a_2^{\dagger })^n\left| 00\right\rangle $. The NOPA
states $\left| {\rm NOPA}\right\rangle $ are the optical analog of the EPR
entangled states in the limit of infinite squeezing. Thus the EPR's argument
can be tested experimentally with the parametric amplifier \cite{Ou,Ou-APB}.
The squeezed-state entanglement of $\left| {\rm NOPA}\right\rangle $ is also
essential in the teleportation of continuous quantum variables \cite{BK}.

Using Eqs. (\ref{s}), (\ref{as}) and (\ref{nopa}) we derive 
\begin{eqnarray}
\left\langle {\cal B}_{CHSH}\right\rangle &=&E(\theta _a,\theta _b)+E(\theta
_a,\theta _{b^{\prime }})  \nonumber  \label{bchsh} \\
&&+E(\theta _{a^{\prime }},\theta _b)-E(\theta _{a^{\prime }},\theta
_{b^{\prime }}),  \label{bchsh}
\end{eqnarray}
where the correlation function 
\begin{eqnarray}
E(\theta _a,\theta _b) &=&\left\langle {\rm NOPA}\right| s_{\theta
_a}^{\left( 1\right) }\otimes s_{\theta _b}^{\left( 2\right) }\left| {\rm %
NOPA}\right\rangle  \nonumber \\
\ &=&\cos \theta _a\cos \theta _b+K\sin \theta _a\sin \theta _b,
\label{corr-NOPA} \\
s_{\theta _a}^{(j)} &\equiv &s_{jz}\cos \theta _a+s_{jx}\sin \theta _a
\label{eth}
\end{eqnarray}
with $K(r)\equiv \tanh (2r)\leq 1$. In deriving Eq. (\ref{bchsh}) we have
set all azimuthal angles to be zero without affecting the following
discussion. Choosing $\theta _a=0$, $\theta _{a^{\prime }}=\pi /2$ and $%
\theta _b=-\theta _{b^{\prime }}$, we have 
\begin{equation}
\left\langle {\cal B}_{CHSH}\right\rangle =2(\cos \theta _b+K\sin \theta _b).
\label{bch}
\end{equation}
For this specific setting the maximum of $\left\langle {\cal B}%
_{CHSH}\right\rangle $ is 
\begin{equation}
\left\langle {\cal B}_{CHSH}\right\rangle _{%
\mathop{\rm max}
}=2\sqrt{1+K^2}  \label{max}
\end{equation}
when $\theta _b=\tan ^{-1}K$. Thus the NOPA states always violate the
Bell-CHSH inequality (\ref{chsh}) provided that $r\neq 0$. Meanwhile, the
degree of quantum nonlocality uncovered here is {\it uniquely} determined by
the squeezing parameter $r$; the parameter $K$ may be reasonably regarded as
a quantitative measure of quantum nonlocality. Compared with Ref. \cite
{Banaszek}, here we do not rely on the phase-space formalism.

The NOPA states $\left| {\rm NOPA}\right\rangle $ can also be written as 
\cite{Banaszek}: 
\begin{equation}
\left| {\rm NOPA}\right\rangle =\sqrt{1-\tanh ^2r}\int dq\int dq^{\prime
}g(q,q^{\prime };\tanh r)|qq^{\prime }\rangle ,  \label{epr}
\end{equation}
where $g\left( q,q^{\prime };x\right) \equiv \frac 1{\sqrt{\pi (1-x^2)}}\exp
\left[ -\frac{q^2+q^{\prime 2}-2qq^{\prime }x}{2(1-x^2)}\right] $ and $%
\left| qq^{\prime }\right\rangle \equiv \left| q\right\rangle _1\otimes
\left| q^{\prime }\right\rangle _2$, with $\left| q\right\rangle $ being the
usual eigenstates of the position operator. Since $\lim_{x\rightarrow
1}g\left( q,q^{\prime };x\right) =\delta (q-q^{\prime })$, one has $%
\lim_{r\rightarrow \infty }\int dq\int dq^{\prime }g(q,q^{\prime };\tanh
r)|qq^{\prime }\rangle =\int dq|qq\rangle =\left| {\rm EPR}\right\rangle $,
which is just the original EPR states. Thus, in the infinite squeezing
limit, $\left| {\rm NOPA}\right\rangle \left| _{r\rightarrow \infty }\right. 
$ become the original, normalized EPR states, for which we have 
\begin{equation}
\left\langle {\cal B}_{CHSH}\right\rangle _{%
\mathop{\rm max}
}=2\sqrt{2}  \label{epr-m}
\end{equation}
by noting $K(r\rightarrow \infty )=1$ and choosing $\theta _a=0$, $\theta
_{a^{\prime }}=\pi /2$ and $\theta _b=-\theta _{b^{\prime }}=\pi /4$ in Eq. (%
\ref{bchsh}). This remarkable result indicates that {\it the normalized
version of the original EPR states can maximally violate the Bell-CHSH
inequality} (\ref{chsh}).

Having shown {\it theoretically} the violation of the Bell-CHSH inequality
by the NOPA states, an important question arise as to what physical
measurements are necessary to test {\it experimentally} quantum mechanics
versus local realism within the present scheme. For this purpose, it is
sufficient to consider how to measure $s_{\theta _a}^{(j)}$ in Eq. (\ref{eth}%
) so that the correlation function $E(\theta _a,\theta _b)$ can be obtained.
Thus in the following we discuss possible schemes of measuring $s_\theta
\equiv s_z\cos \theta +s_x\sin \theta $ for an arbitrary single-mode state
of CVs. Quantum mechanically, $s_\theta $ represents an observable and thus
can be measured in principle. But measuring it in practice is nontrivial.

In Ref. \cite{GCZ}, a scheme is presented to measure an arbitrary motional
observable of a trapped ion. It may be used if one plans to test the present
Bell-CHSH inequality with two trapped ions in entangled motional states.
Here we consider the case where the entangled states of CVs are prepared
within two spatially separated high quality cavities \cite{Jeong}, each of
which resonantly interacts with a sequence of $N$ two-level atoms. It
suffices to consider only one of the cavities characterized by the
annihilation operator $a$ of the cavity field. The atom-cavity interaction
is the usual Jaynes-Cummings Hamiltonian. Assuming the interaction time of
each atom is $t_I$, the unitary evolution operator of the total atom-cavity
system, in the interaction picture, is \cite{complete} 
\begin{equation}
U_N(t_I)=e^{-igt_I(a^{\dagger }\sigma _N+a\sigma _N^{\dagger })}\cdots
e^{-igt_I(a^{\dagger }\sigma _1+a\sigma _1^{\dagger })},  \label{natom}
\end{equation}
where $\sigma _i=\left| g\right\rangle _{ii}\left\langle e\right| $ ($\left|
g\right\rangle _i$ and $\left| e\right\rangle _i$ are the ground and excited
states, respectively, of the $i$th atom) and $g$ is the atom-cavity coupling
strength. Under the condition that trapping states are absent, the property
of ``asymptotic completeness'' can be proved \cite{complete}: 
\begin{equation}
\lim_{N\rightarrow \infty }U_N^{\dagger }(t_I)(A\otimes
I_N)U_N(t_I)=I\otimes M_A,  \label{asym}
\end{equation}
which means that every observable $A$ (e.g., $s_\theta $) of the cavity
field fully develops into the corresponding observable $M_A$ of atoms in the
asymptotic limit. Here $I_N$ is the unit operator in the Hilbert space of
the $N$ atoms. Reversing the interaction time $t_I$ in (\ref{asym}), the
asymptotic completeness should be still valid, and reads 
\begin{eqnarray}
\lim_{N\rightarrow \infty }W_N(t_I)(A\otimes I_N)W_N^{\dagger }(t_I)
&=&I\otimes M_A,  \label{asymt} \\
W_N(t_I) &\equiv &U_N^{\dagger }(-t_I),  \label{uw}
\end{eqnarray}
providing that $A$ and $M_A$ are time-independent. Using the asymptotic
completeness (\ref{asymt}), the expectation value of $A$ with respect to a
cavity field state $\left| f\right\rangle $ is 
\begin{eqnarray}
\left\langle f\right| A\left| f\right\rangle &=&\left\langle N\right|
\left\langle f\right| A\otimes I_N\left| f\right\rangle \left| N\right\rangle
\nonumber \\
\ &=&\lim_{N\rightarrow \infty }\left\langle N\right| \left\langle f\right|
W_N^{\dagger }(I\otimes M_A)W_N\left| f\right\rangle \left| N\right\rangle
\label{faf}
\end{eqnarray}
with $\left| N\right\rangle $ being the initial state of atoms. Equation (%
\ref{faf}) implies that to measure $\left\langle f\right| A\left|
f\right\rangle $, one can send the $N$ atoms across the cavity in sequence;
in the limit $N\rightarrow \infty $, $\left\langle f\right| A\left|
f\right\rangle $ is fully determined by measuring the observable of atoms $%
M_A$ only.

However, it is impossible to handle practically an infinite number of atoms
as above. Fortunately, with a finite number $N$ of atoms, the above strategy
can still yield the desired result with high accuracy, in fact, approaching $%
100\%$ exponentially fast in $N$ \cite{complete}. In this respect, it is
also important to choose $\left| N\right\rangle $ properly to obtain optimal
accuracy. Since states of atoms can be manipulated and measured with current
technology with good accuracy \cite{QIT}, the strategy described here,
though still experimentally challenging, offers a feasible way to measure
the correlation function $E(\theta _a,\theta _b)$ with acceptable accuracy.
But more elegant method of measuring the correlation function $E(\theta
_a,\theta _b)$ is highly desirable.

To summarize, we have defined a new Bell operator for biparty systems with
CVs. In this way Bell's inequalities have been generalized to CV systems. It
is then demonstrated that the NOPA states display quantum nonlocality by
using the Bell operator. In the limiting case of infinite squeezing, the
NOPA states reduce to the original normalized EPR states which are shown to
maximally violate the Bell-CHSH inequality. A strategy to approximately test
the Bell-CHSH inequality has been proposed. The present work reveals a
perfect analogy between the CV systems and the usual spin-$1/2 $ systems.
This fact opens up the possibility that in terms of the parity spin, the CV
systems may be exploited to do quantum information tasks (e.g., quantum
teleportation \cite{PRL-Nature}) as if they were usual qubits. Since the
parity spin operator acts as a collective operator, we expect that our
method might be robust against photon losses. Moreover, the present
formulation enables us to derive all types of Bell's theorem for CV systems.
For instance, the extension of our result to the Greenberger-Horne-Zeilinger
theorem \cite{GHZ} of CVs is also possible and straightforward \cite
{qp0103082}.

We thank K. Banaszek and Z.Y. Jeff Ou for valuable discussions. This work
was supported by the National Natural Science Foundation of China and by the
Chinese Academy of Sciences.

{\it Note added} -- After submission of our work, Rob Clifton brought our
attention to a related paper [H. Halvorson, Lett. Math. Phys. 53, 321 (2000)


\begin{references}
\bibitem{EPR}  A. Einstein, B. Podolsky and N. Rosen, Phys. Rev. {\bf 47},
777 (1935).

\bibitem{Bell}  J. S. Bell, Physics (Long Island) {\bf 1,} 195 (1964).

\bibitem{CHSH}  J. F. Clauser, M. A. Horne, A. Shimony and R. A. Holt, Phys.
Rev. Lett. {\bf 23,} 880 (1969).

\bibitem{Bell-book}  J. S. Bell, {\it Speakable and Unspeakable in Quantum
Mechanics} (Cambridge Univ. Press, Cambridge, 1987).

\bibitem{Bohm}  D. Bohm, {\it Quantum Theory} (Prentice Hall, Englewood
Cliffs, NJ, 1951).

\bibitem{Reid}  M. D. Reid and P. D. Drummond, Phys. Rev. Lett. {\bf 60,}
2731 (1988); M. D. Reid, Phys. Rev. A{\bf \ 40,} 913 (1989).

\bibitem{Walls-book}  D. F. Walls and G. J. Milburn, {\it Quantum Optics}
(Springer-Verlag, Berlin, 1994).

\bibitem{Ou}  Z. Y. Ou, S. F. Pereira, H. J. Kimble, and K. C. Peng, Phys.
Rev. Lett. {\bf 68,} 3663 (1992).

\bibitem{Ou-APB}  Z. Y. Ou, S. F. Pereira, and H. J. Kimble, Appl. Phys. B:
Photophys. Laser Chem. {\bf 55}, 265 (1992).

\bibitem{QIT}  D. Bouwmeester, A. Ekert, and A. Zeilinger (eds), {\it The
Physics of Quantum Information} (Springer-Verlag, Berlin, 2000).

\bibitem{Nature}  C. H. Bennett and D. P. DiVincenzo, Nature (London) {\bf %
404}, 247 (2000).

\bibitem{BK}  L. Vaidman, Phys. Rev. A{\bf \ 49,} 1473 (1994); S. L.
Braunstein and H. J. Kimble, Phys. Rev. Lett. {\bf 80,} 869 (1998); A.
Furusawa {\it et al}., Science {\bf 282,} 706 (1998).

\bibitem{error}  S. L. Braunstein, Phys. Rev. Lett. {\bf 80,} 4084 (1998);
S. Lloyd and J.-J. E. Slotine, {\it ibid}. {\bf 80,} 4088 (1998); S. L.
Braunstein, Nature (London) {\bf 394,} 47 (1998).

\bibitem{computation}  S. Lloyd and S. L. Braunstein, Phys. Rev. Lett. {\bf %
82,} 1784 (1999).

\bibitem{Duan}  L.-M. Duan {\it et al}., Phys. Rev. Lett. {\bf 84,} 4002
(2000).

\bibitem{cloning}  N. J. Cerf, A. Ipe, and X. Rottenberg, Phys. Rev. Lett. 
{\bf 85,} 1754 (2000).

\bibitem{Bell86}  J. S. Bell, Ann. (N.Y.) Acad. Sci. {\bf 480}, 263 (1986).

\bibitem{Banaszek}  K. Banaszek and K. W\'odkiewicz, Phys. Rev. A{\bf \ 58},
4345 (1998); Phys. Rev. Lett. {\bf 82,} 2009 (1999); Acta Phys. Slov. {\bf 49%
}, 491 (1999).

\bibitem{Wigner}  E. P. Wigner, Phys. Rev. {\bf 40,} 749 (1932).

\bibitem{Yurke}  B. Yurke and D. Stoler, Phys. Rev. Lett. {\bf 79,} 4941
(1997).

\bibitem{exp}  A. Kuzmich, I. A. Walmsley, and L. Mandel, Phys. Rev. Lett. 
{\bf 85,} 1349 (2000).

\bibitem{Grangier}  P. Grangier, M. J. Potasek, and B. Yurke, Phys. Rev. A%
{\bf \ 38,} 3132 (1988).

\bibitem{Jeong}  H. Jeong, J. Lee, and M. S. Kim, Phys. Rev. A{\bf \ 61,}
052101 (2000).

\bibitem{Bell-op}  S. L. Braunstein, A. Mann, and M. Revzen, Phys. Rev.
Lett. {\bf 68,} 3259 (1992).

\bibitem{Landau}  L. J. Landau, Phys. Lett. A{\bf \ 120,} 54 (1987).

\bibitem{MAX}  B. S. Cirel'son, Lett. Math. Phys. {\bf 4}, 93 (1980).

\bibitem{Susskind}  L. Susskind and J. Glogower, Physics {\bf 1}, 49 (1964).

\bibitem{GCZ}  S. A. Gardiner, J. I. Cirac, and P. Zoller, Phys. Rev. A {\bf %
55}, 1683 (1997).

\bibitem{complete}  T. Wellens {\it et al}., Phys. Rev. Lett. {\bf 85}, 3361
(2000).

\bibitem{PRL-Nature}  C. H. Bennett {\it et al}., Phys. Rev. Lett. {\bf 70,}
1895 (1993); D. Bouwmeester, J.-W. Pan, K. Mattle, M. Eibl, H. Weinfurter,
and A. Zeilinger, Nature (London) {\bf 390,} 575 (1997).

\bibitem{GHZ}  D. M. Greenberger, M. A. Horne, A. Shimony, and A. Zeilinger,
Am. J. Phys. {\bf 58}, 1131 (1990); J.-W. Pan {\it et al}., Nature (London) 
{\bf 403}, 515 (2000).

\bibitem{qp0103082}  Z. B. Chen and Y. D. Zhang, quant-ph/0103082.
\end{references}
\end{document}